\documentclass[12pt]{article}
\usepackage{scicite}
\usepackage{graphicx}
\usepackage{times}
\usepackage{amsmath}
\usepackage{hyperref}
\usepackage{verbatim}

\newenvironment{sciabstract}{%
\begin{quote} \bf}
{\end{quote}}

\title{Evidence of In-Memory Computing in a Ferrofluid}

\author{Marco Crepaldi$^{1}$, Charanraj Mohan$^{1}$, Erik Garofalo$^{2}$,
\\
Andrew Adamatzky$^{3}$, Konrad Szaciłowski$^{4}$, Alessandro Chiolerio$^{2\ast}$\\
\\
\normalsize{$^{1}$Electronic Design Laboratory,}\\
\normalsize{Istituto Italiano di Tecnologia,16163, Genova, Italy}\\
\normalsize{$^{2}$Bioinspired Soft Robotics,}\\ 
\normalsize{Istituto Italiano di Tecnologia,16163, Genova, Italy}\\
\normalsize{$^{3}$Unconventional Computing Laboratory,}\\
\normalsize{University of West England, BS16 1QY, Bristol, United Kingdom}\\
\normalsize{$^{4}$Academic Centre for Materials and Nanotechnology,}\\ 
\normalsize{AGH University of Science and Technology, 30-059, Kraków, Poland}\\
\\
\normalsize{$^\ast$To whom correspondence should be addressed; E-mail:  alessandro.chiolerio@iit.it}
}
\date{}

\begin{document} 


\baselineskip24pt


\maketitle 


\begin{sciabstract}

  %Currently, electrical computing systems are almost uniquely made of shapes of solid-state matter.
  %However, investigations on colloids featuring functional nanoparticles open a new field of science towards colloidal computers where memory and computing roles can coexist in a liquid aggregation state, regardless of their shape. 
  %Magnetic fluids are applicable in important research fields including energy harvesting, biomedical applications, soft robotics and exploration.
  %Ferrofluids (FFs) are shape reconfigurable and exhibit reversible paramagnetic-to-ferromagnetic transformation.
  Magnetic fluids are excellent candidates for important research fields including energy harvesting, biomedical applications, soft robotics and exploration. However, notwithstanding relevant advancements such as shape reconfigurability, that have been demonstrated, there is no evidence for their computation capability, including the emulation of synaptic functions.
  %, which requires complex non-linear dynamics.
  Here, we experimentally demonstrate that a $\mathrm{Fe}_3 \mathrm{O}_4$ water-based Ferrofluid (FF) can perform electrical analog computing and be programmed using quasi DC signals and read at Radio Frequency (RF) mode. We have observed features in all respects attributable to a memristive behavior, featuring both short and long-term information storage capacity and plasticity. The colloid was capable of classifying digits of a 8\,$\times$\,8 pixel dataset using a custom in-memory signal processing scheme, and through Physical Reservoir Computing (PRC) by training a readout layer. 
  %These findings demonstrate the feasibility of electrical analog computing using a colloid in a liquid aggregation state. 
\end{sciabstract}

% In setting up this template for *Science* papers, we've used both
% the \section* command and the \paragraph* command for topical
% divisions.  Which you use will of course depend on the type of paper
% you're writing.  Review Articles tend to have displayed headings, for
% which \section* is more appropriate; Research Articles, when they have
% formal topical divisions at all, tend to signal them with bold text
% that runs into the paragraph, for which \paragraph* is the right
% choice.  Either way, use the asterisk (*) modifier, as shown, to
\paragraph*{One-Sentence Summary}
A Ferrofluid can store analog information and run electrical in-memory computing using quasi-DC write and RF read modes.

%\section*{Introduction}
\section*{Main Text}
As a part of the largest international effort underway to explore alternative computing methods called \emph{unconventional computing}\cite{bib8,adamatzky2021handbook},
there is a consolidated trend in the research on devices, materials and in natural processes, to find an implicit exhibition of computing features, even beyond solid aggregation state.
The idea of computing with liquids attracted engineers and mathematicians since the early 1900s~\cite{emch1901two}, but later prototypes of liquid computers were mostly based on hydraulic, reaction-diffusion, and fluidic principles \cite{adamatzky2019brief}.
%included hydraulic integrators, a monetary national income analog computer, fluid mappers, fluid logic, reaction-diffusion computers and fluid maze solvers .
Only recently, liquid and colloidal systems have been subject to attention for mimicking the ions moving in the human brain through embedding aqueous solutions in gel or solid-state scaffolds \cite{noushin}.
Being applicable to soft robotics \cite{bib76}, energy harvesting \cite{bib77}, and computation in general \cite{bib78}, 
magnetic fluids are always of great research interest (Sec.~S1). Particularly, ferrofluids (FFs) are mixtures in which nanometric-size dispersed insoluble particles are suspended throughout a solvent, the particles being typically superparamagnetic, giving rise to interesting collective behaviour. 
A potential of FF in computing, massive-parallel information processing, sensing, and energy harvesting regardless of their shape has not been addressed before, notwithstanding first reports on their shape reconfiguration are already available \cite{shape1}. Here, we demonstrate that a volume of a superparamagnetic FF can be interchangeably assigned to memory and computing roles. This property is consistent with the rising paradigm of in-memory computing, which aims at mitigating processor-memory data transfer bottleneck by embedding computation in memory \cite{ielmini2018memory,verma2019memory,le2018mixed,sebastian2020memory}. Furthermore, we demonstrate FF computation using the concept of Reservoir Computing (RC), a paradigm that takes advantage from system dynamics (spontaneous or excited from external sources) for advanced information processing. Reversibility, fading memory, nonlinearity of electrical response and structural stochasticity are usually considered as prerequisites for any physical implementation of RC concepts \cite{reservoir}, and most solid-state memristors fulfil these requirements \cite{reservoir2}.  
%Here, for the very first time we have observed and experimentally demonstrated electrical computing capabilities in a superparamagnetic water-based FF.
%, and features attributable to short and long term plasticity. 

%Conventional computer architectures have separate locations for data storage and computing, thus resources are wasted on information transfers between a memory and a processor. 

%\section*{Experimental Set-Up and First Proof}

\begin{figure}[h]%
\centering
\includegraphics[width=0.85\textwidth]{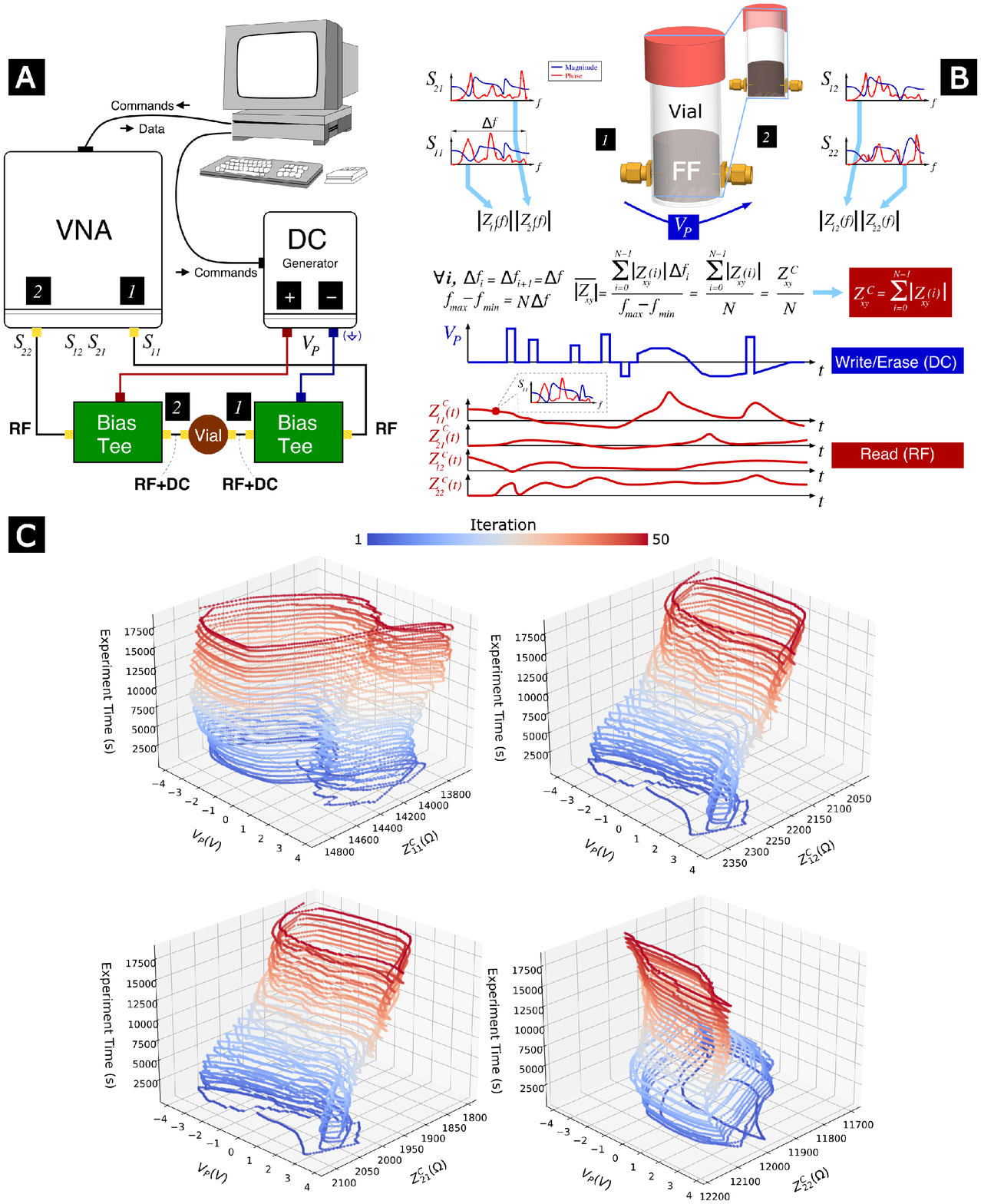}
\caption{\textbf{A} Experimental set-up. \textbf{B} Measurement concept and relevant parameters. The colloid is programmed using a quasi-DC voltage and its internal status is read through its distributed impedance values $Z_{11}^C$, $Z_{12}^C$, $Z_{21}^C$ and $Z_{22}^C$, that correspond to the sum of the samples obtained throughout the measurement bandwidth of the VNA (10\,MHz--6\,GHz). \textbf{C} Hysteresis loops as a function of experiment elapsed time obtained from an initial impedance multi-point (comprising $Z_{11}$, $Z_{12}$, $Z_{21}$ and $Z_{22}$) obtained by applying a -3.8\,V--3.8\,V voltage sweep.}\label{fig1}
\end{figure}

Our experimental setup shown in Fig.~\ref{fig1}A comprises a FF sample (the reservoir) connected to a two-port Vector Network Analyzer (VNA), a DC bias generator (where the negative terminal is internally connected to ground) and two bias tee circuits to decouple RF and DC signals (details in Materials and Methods). 
Both the DC bias generator and the VNA are connected to a Personal Computer to implement measurement scripts (Sec.~S2), that is, applying a DC voltage across the reservoir and reading the impedance through its S-parameters, that can be always converted to impedances as a function of frequency.
As shown in Fig.~\ref{fig1}B, the FF is stimulated as follows: a quasi-DC voltage $V_P$ (both positive and negative) is applied to the system to program/write it, and its internal status (read) is acquired in RF mode using the magnitude impedance parameters. Since the observed S-parameters and consequently impedance magnitude variations are small (Sec.~S3), the sum of the numerical impedance values over all scanned frequencies $Z_{xy}^C$ can be used profitably as an indicator (see the formulas in the figure). This way the S-parameters are collapsed to a single number for each measurement point, reducing data volume. Consequently, each reading measurement is an ensemble of four real numbers $Z_{11}^C$, $Z_{12}^C$, $Z_{21}^C$ and $Z_{22}^C$, all a function of time $t$, e.g., for port one, $Z_{11}^C\equiv Z_{11}^C(t)$. 

Hysteresis, a fingerprint of memristance \cite{chua1}, is a necessary condition for neuromorphic computation \cite{bib72}.
Fig.~\ref{fig1}C shows hysteresis loops obtained with a voltage sweep from -3.8\,V to 3.8\,V with steps of 0.1\,V each lasting 1\,s, repeated for 50 times. At the beginning of the test ($t$\,=\,0\,s) we started with $V_P$\,=\,-0.85\,V and impedance values were $Z_{11}^C(0)$\,=\,14312\,$\Omega$, $Z_{12}^C(0)$\,=\,2320\,$\Omega$, $Z_{21}^C(0)$\,=\,2060\,$\Omega$ and $Z_{22}^C(0)$\,=\,11795\,$\Omega$. 
The pinched hysteresis of $Z_{11}^C$ shrinks for positive $V_P$ as the number of iterations increases, even with such zero average excitation. The hysteresis of $Z_{22}^C$ shrinks throughout the whole $V_P$ range while for $Z_{12}^C$ and $Z_{21}^C$ we observe the opposite phenomenon. On the one hand, the results indicate that assuming a given DC stimulus, its effect on the impedance variation is not constant and varies over time. On the other hand, this feature indicates a long-term adjustment of the material towards an \emph{equilibrium} condition, that can be interpreted as the feature of memorizing the previous DC bias history. Furthermore, due to fluidity of the material, this memory will be naturally fading, which is another important prerequisite for an efficient and universal RC system \cite{memory1}.

%\section{Analog Memory}

\begin{figure}[h]%
\centering
\includegraphics[width=0.9\textwidth]{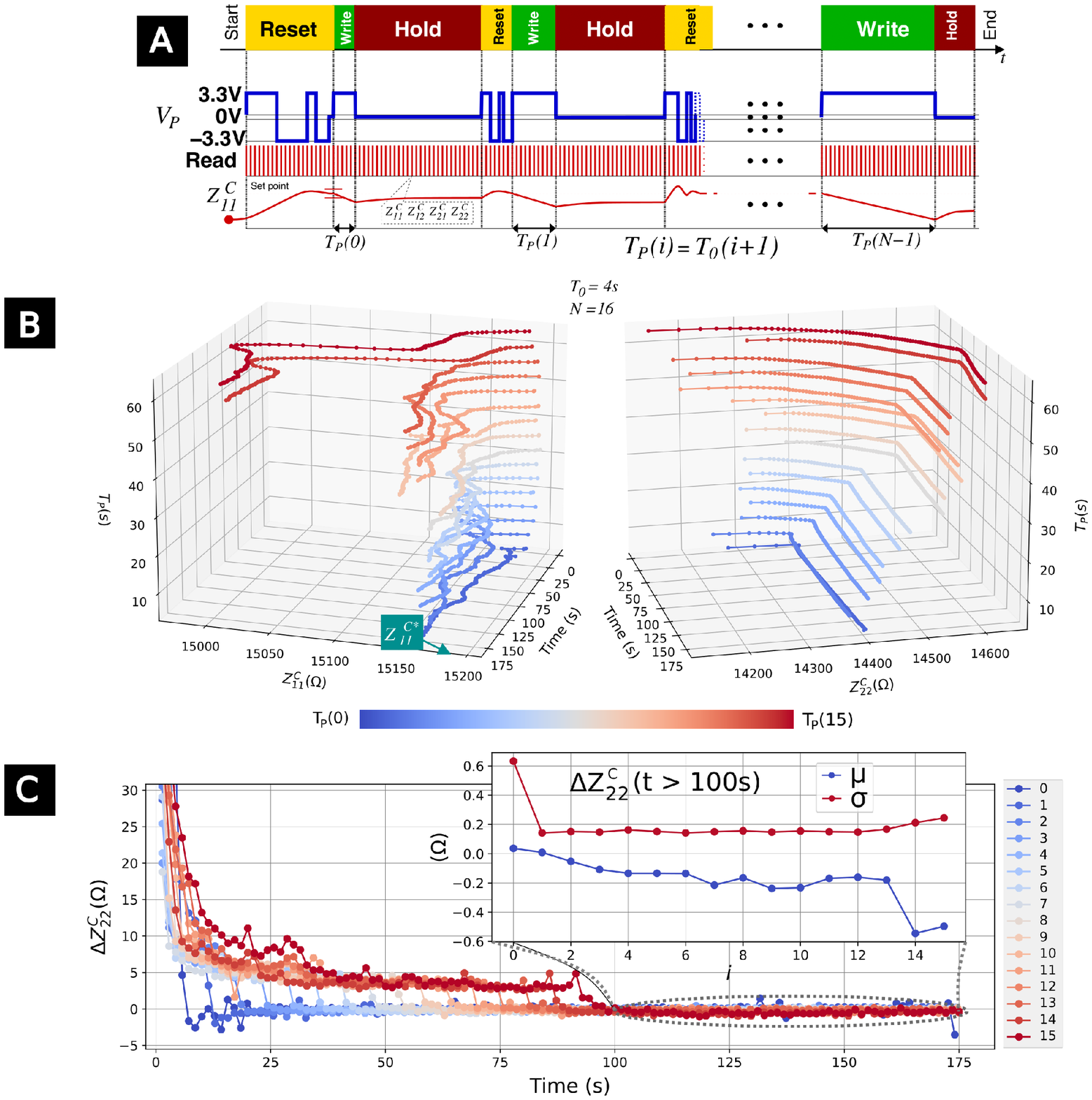}
\caption{\textbf{A} Long-term memory stimulus scheme and \textbf{B} results obtained by applying a positive pulse to the material of value 3.3\,V with different duration $T_P(i)$, and by restoring the initial impedance value $\mathrm{Z}_{11}^{C*}$\,=\,14338\,$\Omega$ for each test using a closed control loop. \textbf{C} Variation of $\mathrm{Z}_{22}^{C}$ and corresponding mean and variance during the {\tt Hold} phase for all information values. }\label{fig3}
\end{figure}

The liquid can be used to store information in the form of a particular impedance evolution at a given port. To pose a parallelism with biological neurons we can refer to a \emph{long-term plasticity} feature.
Fig.~\ref{fig3}A shows the stimulus scheme used to demonstrate storage capacity for $N$ information values -- in this specific test $N$\,=\,16. The test comprises repeated {\tt Reset}, {\tt Write} and {\tt Hold} phases, where {\tt Reset} implements a control loop on $\mathrm{Z}_{11}^C$ to reset its value to $\mathrm{Z}_{11}^{C*}$. 
The results in Fig.~\ref{fig3}B show that $\mathrm{Z}_{11}^{C*}$ is correctly reached for each iteration, and notwithstanding impedance control is implemented at port one, $\mathrm{Z}_{22}^C$ evolves towards well defined impedance values, that are a function of the applied pulse duration $T_P(i)$. Interestingly, the $\mathrm{Z}_{22}^C$ values do not reset at the beginning of each {\tt Write} phase. The small variation of the $\Delta Z_{22}^C$ values and the associated uniform distribution parameters during {\tt Hold} of Fig.~\ref{fig3}C, suggests that the colloid can be used as a high resolution short-term memory. In general, as $T_P(i)$ can be controlled with the power of continuum, analog information storage can be implemented and information can be stored even for a longer duration (Sec.~S4). 

%\section{Computing}

%\subsection{Pattern Classification}
%\label{secClassification}
\begin{figure}[h]%
\centering
\includegraphics[width=0.9\textwidth]{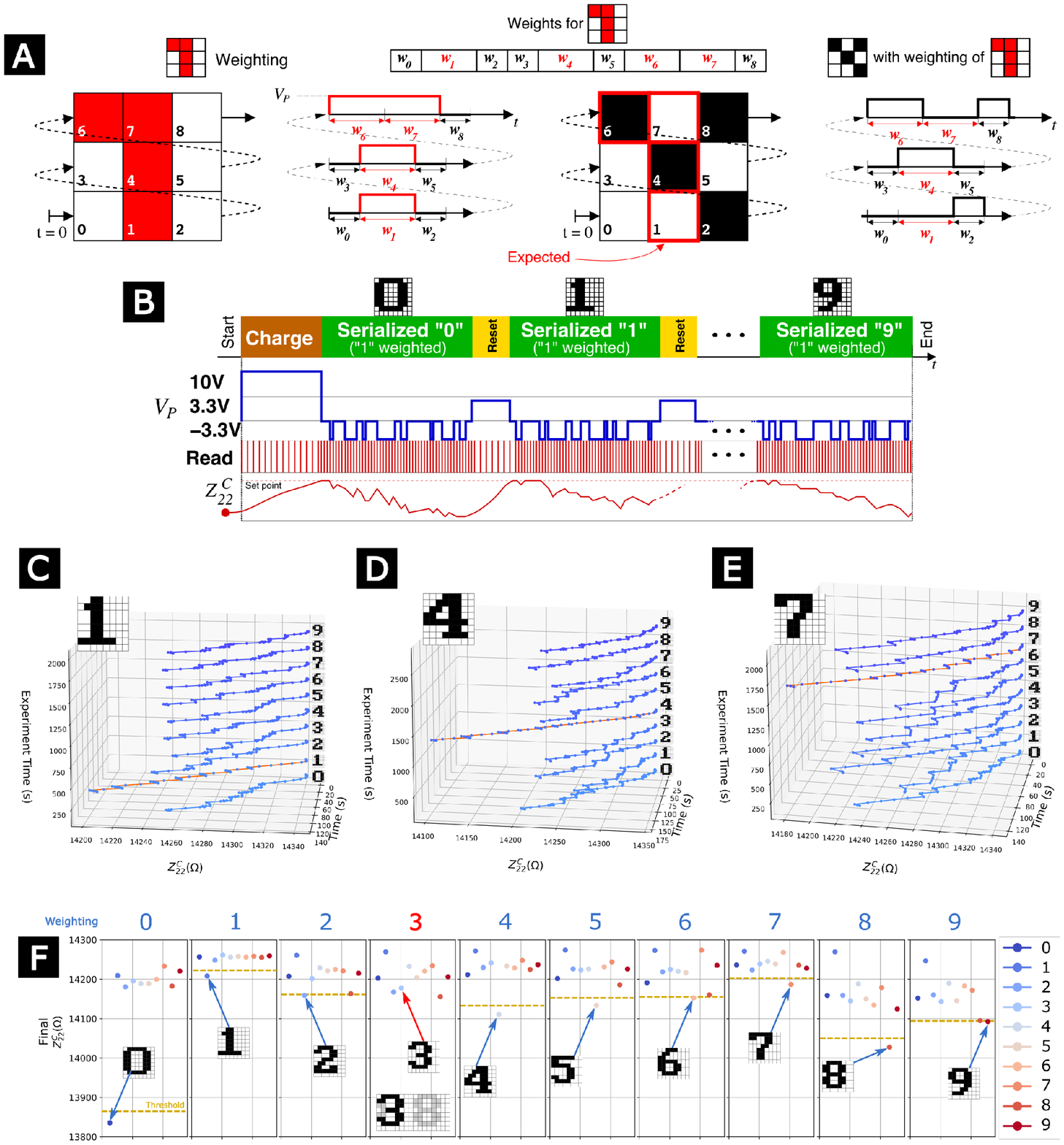}
\caption{\textbf{A} Weighting used for in-memory digit filtering, based on a simple elongation of the expected black pixels pulses. Each digit is serialized from bottom-left to top-right, sequentially line by line.
\textbf{B} Stimulus scheme for classification, with reset control on $Z_{22}^C$ after each serialized digit (exemplified here for {\tt 1} weighting). \textbf{C, D, E} Measured impedance variation for all digits for three examples weighted sequences, {\tt 1}, {\tt 4} and {\tt 7}. Weighting a particular digit leads to a lowering of its impedance $Z_{22}^C$ compared to the others (orange paths).
\textbf{F} Final $Z_{22}^C$ for all weighting sequences. Classification for digit {\tt 3} fails due to overlapping with {\tt 8}.}\label{fig5}
\end{figure}

By extending the memory stimulus scheme of Fig.~\ref{fig3}A, we could demonstrate in-memory digit classification. To this end, 
we prepared a dataset consisting of the ten digits {\tt 0}--{\tt 9}, in 8\,$\times$\,8 matrices of pixels (Sec.~S5) that we have serialized as shown in the scheme of Fig.~\ref{fig5}A. Data serialization has been demonstrated to be an efficient approach towards neuromorphic data processing with very minimal computational resources [e.g. single artificial neurons \cite{molecules, serial2}].
Each pixel, besides its value 0--1 that can be mapped to a voltage level $V_{P}$, can be attributed a weight in terms of pulse duration $w_i$ and, in general, an offset (in terms of additive voltage). It is therefore possible to build up sequences with higher or lower sensitivity to a particular digit or selectively filter particular pixel values. 
Here, we do not bias the colloid in the condition of having a pinched hysteresis, so to obtain a monotonic decrease of impedance during the tests, and therefore enable a direct comparison of the final value for all digits after the application of the sequences. 
By assuming instead that the system is in the conditions of a pinched hysteresis, the FF can progressively adapt to implement a learning mechanism by providing a particular non-zero offset weighting (Sec.~S6).
% weight and bias sequence for a digit that runs across the pinched hysteresis by a larger magnitude compared to others 

Fig.~\ref{fig5}B shows the stimulus scheme of the pattern classification test, which exploits in-memory computing features. Similarly to the memorization experiments here we apply a reset control on $Z_{22}^C$, towards the impedance set-point $Z_{22}^C$\,=\,14338 \,$\Omega$, using an initial {\tt Charge} phase at 10\,V. Each pixel is associated to a -3.3\,V or 0\,V voltage (black or white) for a given weight duration, with zero offset. After verifying differentiation (Sec.~S5), we provide the weighted sequences so that the pulse durations of the expected black pixels are longer compared to the others. We apply sequentially all serialized pixel matrices from {\tt 0} to {\tt 9}, using all the weighted sequences for each digit. Fig.~\ref{fig5}C--E show the measurement results assuming 4.5 and 0.25\,s weights (black and white, respectively), for three sample digits, {\tt 1}, {\tt 4} and {\tt 7}. Results show that with this in-memory computing scheme $Z_{22}^C$ decreases more considerably in case the weighted sequence matches the digit.
The final decision can then be achieved by using a simple threshold on the $Z_{22}^C$ value, which depends on the digit to be detected, or alternatively, by indexing the digit that leads to the lowest impedance after all of them are serialized.
This particular scheme works except for {\tt 3} which is a subset of {\tt 8} (Fig.~\ref{fig5}F), thus leading to 90\% accuracy. 
As an effect of long-term plasticity, we have observed also that if the above test is repeated for days without interruption, the impedance dynamics shrinks, irrespective of the digit (see Sec.~S7). The behaviour of the liquid, however, is reversible and impedance dynamics can be restored. 
%by applying a fixed 'reset' stimulation of -10\,V or waiting for a sufficiently long time.

%\subsection{Physical Reservoir Computing}

\begin{figure}[h]%
\centering
\includegraphics[width=0.9\textwidth]{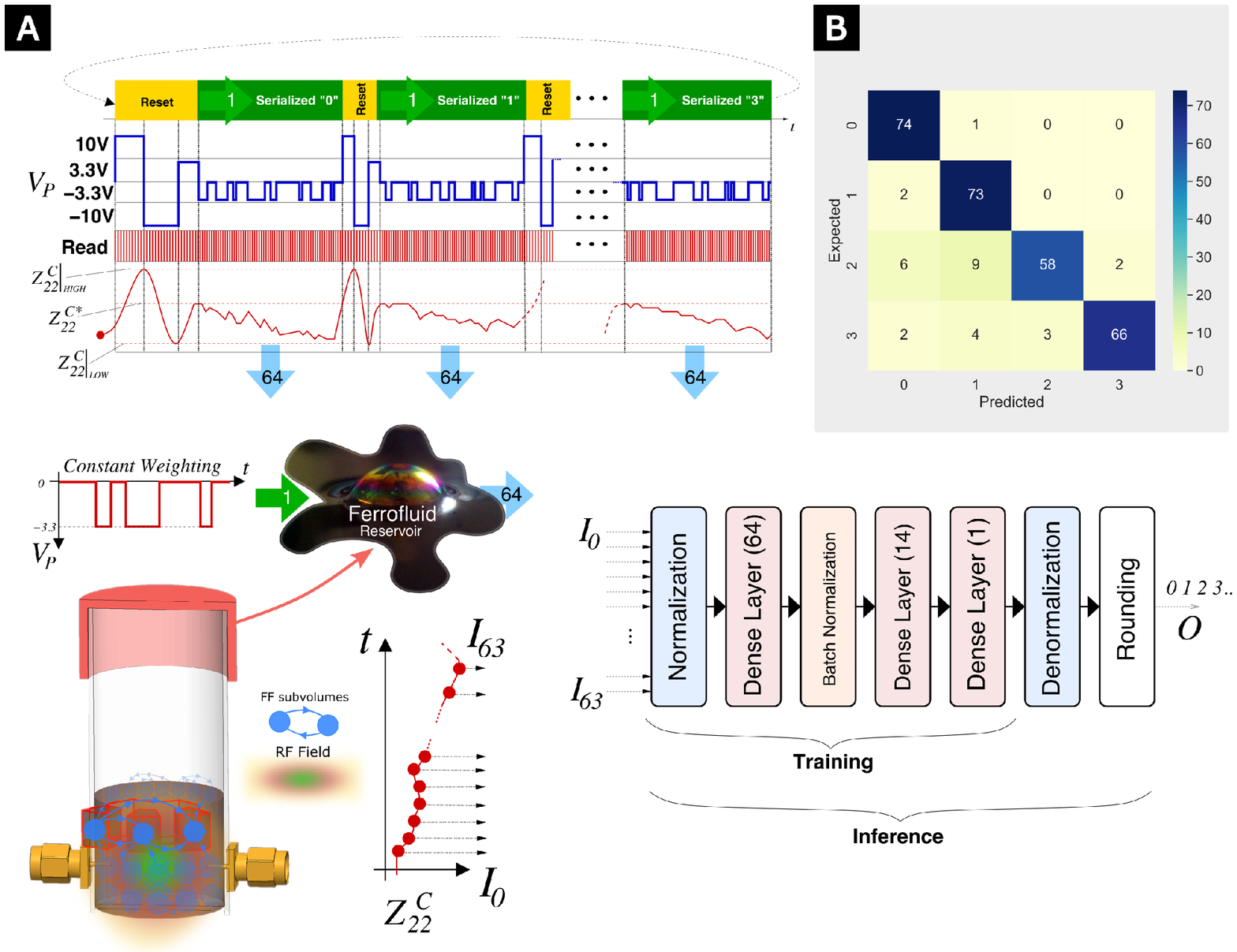}
\caption{\textbf{A} Stimulus scheme used for both training and inference during PRC tests with parallelized outputs for the readout NN layer (each one identifying the effect of a pixel value), detail on the NN layer and conceptual liquid reservoir. \textbf{B} Confusion map of a real-time classification test of the digits {\tt 0} to {\tt 3} using the trained NN.}\label{fig6}
\end{figure}

Similarly to memristors (differences outlined in Sec.~S8), to further demonstrate the computation capability of the FF we have implemented PRC using an ad-hoc readout layer. 
The FF exhibits chaotic nature (resulting, \emph{inter alia} from Brownian motions as well as from the surfactant molecules featuring electrical polarizability), and within its deterministic features, it presents a strong sensitivity to initial electrical conditions (Sec.~S9), while it can provide both fading memory and long-term plasticity (for instance see the plots in Fig.~S3). 
RC is typically implemented taking advantage of a physical reservoir short-term memory \cite{bib67}. However, within the time frame of a digit classification, our findings show that the dynamics of the FF tend in the long-term to reduce if a trivial reset condition is used (Sec.~S7). Moreover, besides sensitivity to initial conditions, the FF exhibits chaotic non-equilibrium at repeated impedance sets (Sec.~S10).
Such features can make PRC unfeasible and it is thus necessary to 
%exploit only its fading memory effect,
avoid changes in dynamical regime during both training and inference. To mitigate these variations, we have designed a particular 'reset' sequence that maintains the dynamical features of the material consistent. It consists of the application of a high voltage towards two impedance points, that are higher and lower compared to the initial impedance $Z_{22}^{C*}$ used to run active computation, respectively. 
Fig.~\ref{fig6}A shows the measurement scheme and the reset sequence. In our tests $Z_{22}^C\rvert_\mathrm{LOW}$\,=\,16350\,$\Omega$, $Z_{22}^C\rvert_\mathrm{HIGH}$\,=\,16450\,$\Omega$ and $Z_{22}^{C*}$\,=\,16400\,$\Omega$.

We used constant weighting to serialize the 64 pixels of the digit matrices (each pixel lasts 2\,s, with -3.3\,V for {\tt 1} and 0\,V for {\tt 0}) and we have trained a Neural Network (NN) layer to classify four digits {\tt 0}--{\tt 3} (rationale and dataset in Sec.~S11). The NN comprises a first block which normalizes in parallel the 64 impedance values in the range 0--1 to get rid of residual dynamical variations due to the FF chaotic nature. The input layer is made of a 64 {\tt Dense} model, followed by a {\tt Batch Normalization} block (that helps back-propagation convergence), another 14 elements {\tt Dense} model, and finally a single {\tt Dense} neuron. Inference is achieved in real-time using the trained NN on new measurement data from the FF consisting on 64 impedance values $Z_{22}^C$. Fig.~\ref{fig6}B shows the confusion map, after detecting the four digits with the pre-trained NN, achieving an accuracy of 90.6\%.

%\section{Conclusion}\label{sec13}

In conclusion, we have demonstrated the first ever evidence of a FF in-memory computing device. A FF can implement complex calculations both with custom in-memory computing schemes, and PRC, thus widening its spectrum of features. Besides extending the possibilities of already existing applications, these findings make solutions featuring unprecedented plasticity, fault-tolerance and resilience towards extreme environments a plausible reality, thanks to FFs amorphous nature.

% Your references go at the end of the main text, and before the
% figures.  For this document we've used BibTeX, the .bib file
% scibib.bib, and the .bst file Science.bst.  The package scicite.sty
% was included to format the reference numbers according to *Science*
% style.

%BibTeX users: After compilation, comment out the following two lines and paste in
% the generated .bbl file. 

%\bibliography{scibib,liquidcomp}

%\bibliographystyle{Science}

%TC:ignore
\paragraph*{Funding}
This project has received funding from the European Union’s Horizon 2020 research and innovation programme FET OPEN ``Challenging current thinking” under grant agreement No 964388.

\paragraph*{Author Contributions}
M.C. devised and prepared the set-up, wrote the software, devised stimuli and ran measurements, analyzed the results, devised and trained the learning network, prepared the manuscript. A.C., conceptualized and devised the set-up, analyzed the results, prepared the manuscript and correspondence. C.M., E.G., A.A. and K.S. prepared the manuscript, analyzed the results and contributed.

\paragraph*{Competing Interests}

The authors declare they have no conflict of interest or competing interests.

\paragraph*{Data and Materials Availability}
Data is available upon reasonable request to the corresponding author.
\section*{Acknowledgments}
We thank Davide Dellepiane, Electronic Design Laboratory, for the assembly of the vial, and Alessandro Barcellona for the support in the design of the bias tee electronics. We thank Diego Torazza, Robotics Brain and Cognitive Sciences dept., for the 3D rendering of the vial.

%Here you should list the contents of your Supplementary Materials -- below is an example. 
%You should include a list of Supplementary figures, Tables, and any references that appear only in the SM. 
%Note that the reference numbering continues from the main text to the SM.
% In the example below, Refs. 4-10 were cited only in the SM.     
\section*{Supplementary Materials}
Materials and Methods\\
Supplementary Text (Sec.~S1 to S11)\\
Supplementary Fig.~S1 to S8\\
%TC:endignore

%\paragraph*{Word Count}
%\quickwordcount{ferrofluid-electrical-computing} including abstract, main text, references and figure captions.
%\detailtexcount{ferrofluid-electrical-computing}
% For your review copy (i.e., the file you initially send in for
% evaluation), you can use the {figure} environment and the
% \includegraphics command to stream your figures into the text, placing
% all figures at the end.  For the final, revised manuscript for
% acceptance and production, however, PostScript or other graphics
% should not be streamed into your compliled file.  Instead, set
% captions as simple paragraphs (with a \noindent tag), setting them
% off from the rest of the text with a \clearpage as shown  below, and
% submit figures as separate files according to the Art Department's
% instructions.

\end{document}

\baselineskip24pt

\title{Supplementary Materials}

\begin{Large}
\begin{center}
Evidence of In-Memory Computing in a Ferrofluid
\end{center}
\end{Large}

\bigskip
\begin{large}
\begin{center}

\textbf{Supplementary Materials}

\end{center}
\end{large}
\bigskip

\noindent\textbf{This PDF file includes:}

\indent Material and Methods

\indent Supplementary Text (Sec.~S1 to S11)

\indent Supplementary Figures (Fig.~S1 to S8) 

\bigskip

\paragraph*{\large{Materials and Methods}}\label{sec11}

\subparagraph{Conditions}
All measurements are performed in an electronic laboratory environment (Electronic Design Laboratory, Via Enrico Melen 83, Genova, Floor 7), at room temperature that ranges 22--25\,$^\circ$\,C, and they are executed at night time to avoid possible vibrations that may occur in the laboratory during normal working hours.

\subparagraph{Ferrofluid}
We have used an EMG601P ferrofluid, FerroTec, Lot Number U021920A \cite{emg601p}. The quantity of liquid used is 5\,ml, that has been released in the vial using a pipette.

\subparagraph{Vial}
The vial is made of an inert Acrylonitrile Butadiene Styrene (ABS) material while the electrical contacts, directly in contact with the liquid, are based on feed-lines of gold plated RF connectors, therefore not contributing to any chemical reaction.
The vial (3\,cm diameter) has been prepared to host RF SMA connectors (Wurth Elektronik, 1.6\,mm straight PCB, Manufacturer number 60314202124525) feed lines. The ground pins of the connectors have been cut to expose only the feed line of the connector. The vial has been drilled to host symmetrically the two feed lines. Finally, the connectors have been threaded in the drilled surface of the vial, and fixed in place using a rubber band and hot glue. In order not to let the liquid evaporate, the vial needs to be closed using the supplied cap.

We have specifically chosen not to implement RF shielding on the vial to avoid undercut propagation modes, therefore permitting the observation of the phenomena without constraints.

\subparagraph{Vector Network Analyzer}
A PicoVNA 106 (300\,kHz--6\,GHz), Pico Technology, UK, has been used to read out the status of the material, using its built-in Dynamic Link Libraries (DLL) under Microsoft Windows 7. In our experiments, the RF power used to perform the frequency sweep by the VNA is 0\,dBm, and we have not observed any significant impact of such a signal on the internal status evolution of the liquid. The number of measurement points was 101.

\subparagraph{Bias Tee}
The two bias tees used are commercial TCBT-14+, Mini Circuits (10\,MHz--10\,GHz), that have been soldered on two custom PCBs designed to be mounted on the RF mini enclosure RF-ENCL-MINI-NF-01, Gequipment.

\subparagraph{DC Generator}
We have implemented the DC generator using a Micropython Board V1.1, connected, through its two available DAC converters to an evaluation board of a Maxim OpAmp with an internal charge pump (MAX 4426T EVAL KIT) to generate, starting from a single 12\,V supply, a $\pm$\,10\,V DC signal using both INAP and INAM terminals. 
The DACs of the Micropython board have been set through the internal firmware to a high current drive. The Micropython board implements a Virtual COM Port (VCP) interface to the Personal Computer so that the measurement program can set the output DC voltage on-demand by asynchronously sending commands to the module. The DC generator accepts an external power supply (GBC 34.0106.10, 18.5\,W, 0.8\,A at 12\,V) that generates the 12\,V supply required for the OpAmp to operate.

\subparagraph{Measurement Software}
The measurement software runs on a Windows 7 Virtual Machine, installed on a CentOS 7 control domain. Both VNA and DC Generator are connected to the PC using USB cables.

To reproduce the measurements presented in the manuscript, it is sufficient to write a program that coordinates both VNA and DC generator to read out the S-parameters from the liquid and set the DC bias point. In this work, however, we have designed a specific Python scripting language that executes and compiles specific experiment files. 

\subparagraph{Impedance Parameters Calculation}
To calculate the impedance parameters starting from the S-parameters we have used the following equations (\ref{eq1}--\ref{eq4}),
\begin{align}
Z_{11} &= \frac{(1+S_{11})(1-S_{22})+S_{21}S_{12}}{\Delta_S}Z_0,\label{eq1}\\
Z_{12} &= \frac{2S_{12}}{\Delta_S}Z_0,\label{eq2}\\
Z_{21} &= \frac{2S_{21}}{\Delta_S}Z_0,\label{eq3}\\
Z_{22} &= \frac{(1-S_{11})(1+S_{22})+S_{21}S_{12}}{\Delta_S}Z_0\label{eq4},
\end{align}
where $\Delta_S = (1-S_{11})(1-S_{22})-S_{21}S_{12}$. The above equations output are the impedance complex numbers values over frequency, from which magnitude values can be extracted. In our measurement system, these calculations are computed during the S-parameters measurements, but they can be calculated offline. We have assumed $Z_0$\,=\,50\,$\Omega$. In our tests, we have not eliminated the contribution of the vial, and we have considered its full impedance contribution including the colloid. 

\subparagraph{PRC Readout Neural Network}
The neural network has been implemented in\linebreak {\tt tensorflow}. Training is achieved using 50 sequences of digits {\tt 0}--{\tt 3} obtained using the same measurement conditions of the other tests (see Sec.~S11).
The optimizer used for training is {\tt Adam} and all the layers have a sigmoid activation function. The 'reset' sequence is not considered in the training, and only the 64 impedance values resulting from serialization are used.
During the real-time inference the measurement system streams the data once all the pixels are serialized, i.e. when a single digit iteration is finished, though the User Datagram Protocol (UDP) over an Ethernet physical layer. The inference is run on a PC where the trained model is loaded and the UDP packets are received from the network.

\subparagraph{Measurements Processing and Graphs}
All the results of this manuscript (except from the PicoVNA 106 that comes with its proprietary DLL and the vial that has been rendered using PTC Creo), have been obtained using open source software, that is Python, Inkscape \linebreak ({\small\tt https://inkscape.org}), Xfig ({\small\tt http://mcj.sourceforge.net}), \linebreak TexMaker ({\small\tt https://www.xm1math.net/texmaker/}) and Gimp \linebreak ({\small\tt https://www.gimp.org}). The files generated by the measurement system have been processed, collated, and organized for plotting and visualization using custom\linebreak {\tt matplotlib} Python utilities.
\newpage

%\section{Extended Introduction}
%\label{Sec1}

%\subsection{Unconventional Computing Examples}

%Some specific unconventional computing examples are DNA computing \cite{bib9}, linear programming machines \cite{bib10}, quantum computing \cite{bib11}, bubble soap \cite{bib12}, optical devices \cite{bib13}, spaghetti sort \cite{bib14}, rainbow sort \cite{bib15}, smart glass \cite{bib16}, protein folding machine \cite{bib17}, and bead sort \cite{bib18}. 

%\subsection{Typical Applications of Ferrofluids}
\begin{large}
\noindent \textbf{Supplementary Text}
\end{large}
\bigskip

\section{Typical Applications of Magnetic Fluids and Ferrofluids}
\label{Sec1}
Magnetic fluids in general are of great interest in biomedical applications e.g., cell separation \cite{bib53}, tumor hyperthermia \cite{bib54}, cancer diagnosis \cite{bib55}, magnetorelaxometry \cite{bib56}, drug delivery \cite{bib57} and thermal energy conversion \cite{bib79}.

Some noteworthy applications of ferrofluids are magnetic seals for pumps and mixers \cite{bib59}, inertial and viscous damping for loudspeakers and stepper motors \cite{bib60}, bearings \cite{bib61}, lubricants \cite{bib62}, heat transfer media \cite{bib63}, and soft robots \cite{bib64,bib65,bib66,bib64c}. 

\newpage

\section{Measurement Software}
\label{Sec3}

We have designed a specific scripting language based on Python to support the experiments through script files and to interface the low-level drivers of the VNA and the DC generator. It comprises a specific software interpreter, capable of compiling the scripts and partially executing them in advance to check their correctness. The possibility of controlling both VNA and DC generator through a PC opens the way to the implementation of programmable experiments, automatic result storage, real-time computation of quantities of interest from the S-parameters, and implementation of control loops to investigate algorithms, for instance, to bring impedance to known set points. 

The experiment files include specific commands to set the DC bias, read the data from the VNA and give the possibility of declaring variables, calculating quantities on the fly based on the measurement output, and thus implementing control loops thanks to the possibility of including selections. The measurement program also takes care of saving the measurement data on the PC and permits customization in the writing of Comma-Separated Values (CSV) files by including not only directly measured S-parameters but also any of the variables used during the execution of the experiment. This approach has the power of enabling the scheduling of specific scripts and deep customizations to enable batch measurements and implementing complex calculations. For instance, the herein referenced impedance values $Z_{11}^C$--$Z_{22}^C$ are calculated by the measurement program immediately after the acquisition of the S-parameters from the VNA. 
\newpage

\section{RF Impedance Controllability}
\begin{figure}[hbt!]%
\centering
\includegraphics[width=0.9\textwidth]{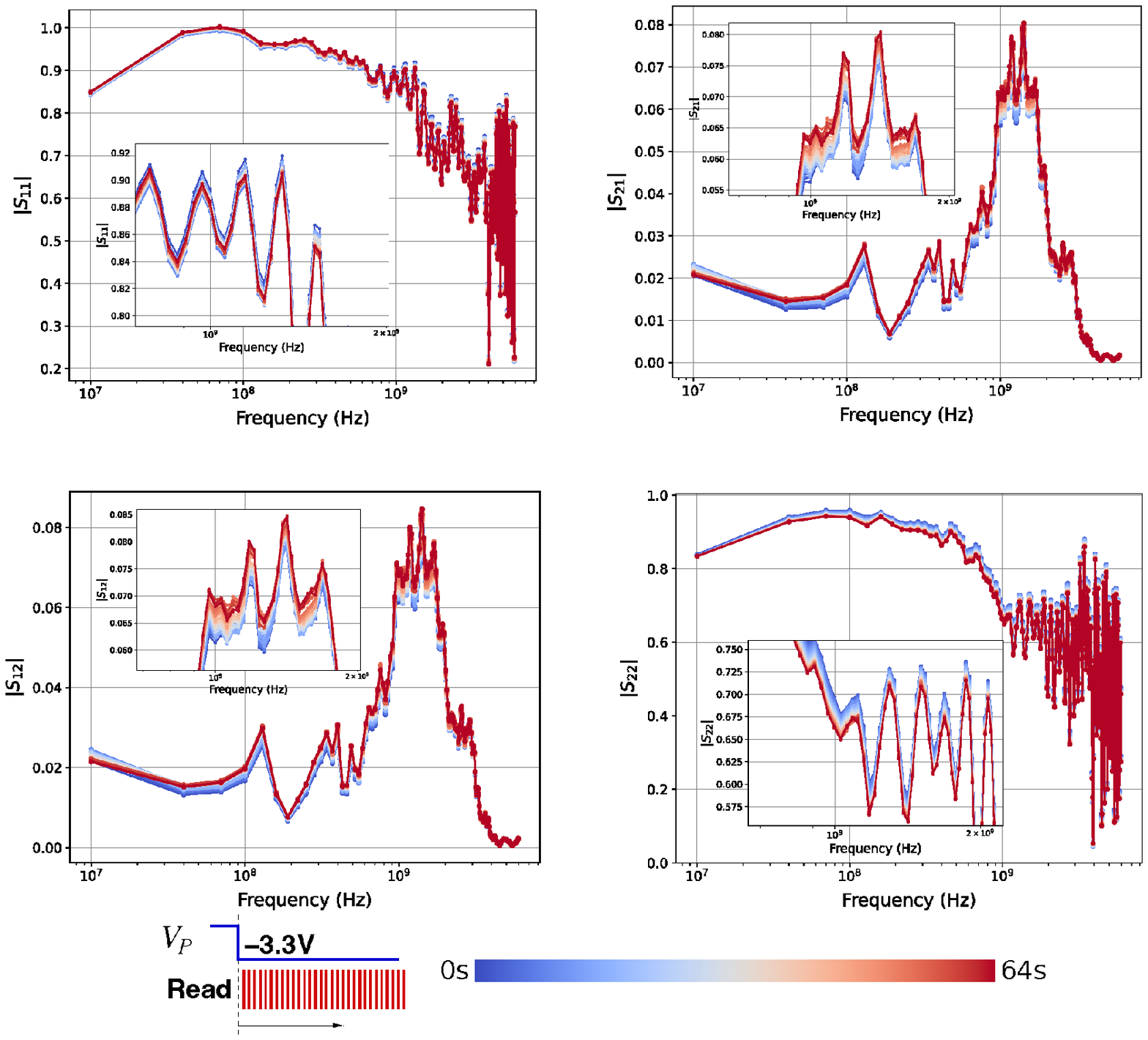}
\caption{Measured S-parameters of the FF with a -3.3\,V bias applied to the material, from 0 to 64\,s, one measurement per second. This measurement demonstrates the DC controllability of the internal particle collective status of the FF, and the possibility of reading it out in RF mode.}\label{figsc}
\end{figure}
\FloatBarrier

Fig.~\ref{figsc} demonstrates the controllability of the liquid RF impedance through the application of a DC bias $V_P$ and shows its consequent S-parameters evolution.
Here, we have applied a constant -3.3\,V DC bias to the material and we have run measurements using the VNA at every second, for 60\,s overall. The graphs show the evolution of the parameters over time, while the negative DC bias is maintained. In general, although $\rvert S_{12}\rvert$ and $\rvert S_{21}\rvert$ slightly differ 
(non-ideal reciprocal behaviour)
%due to a non perfect symmetry of the vial),
different trends occur in the curves. For this experiment, $\rvert S_{22}\rvert$ decreases in the full 10\,MHz--6\,GHz bandwidth, while $\rvert S_{11}\rvert$ for a particular range (see inset) has a different trend. Similar results can be obtained by applying a positive DC voltage. The material behaviour strongly depends on its impedance value at the beginning of the experiment, and the status of the liquid is inherently encoded in the \emph{unbalancing} of all the four parameters. Moreover, the evolution of its internal status is in general reversible by applying a signal with opposite sign. Higher variations in the S-parameters are achievable by applying a larger voltage (up to $\pm$\,10\,V for our set-up), however, we have decided to keep limits at a maximum of $\pm$\,3.3\,V during computation (except for hard resets), for compatibility with the standard voltage levels used in consumer electronic components. 
\newpage

\section{Enhanced Information Storage}
\label{Sec4}
\begin{figure}[hbt!]%
\centering
\includegraphics[width=0.9\textwidth]{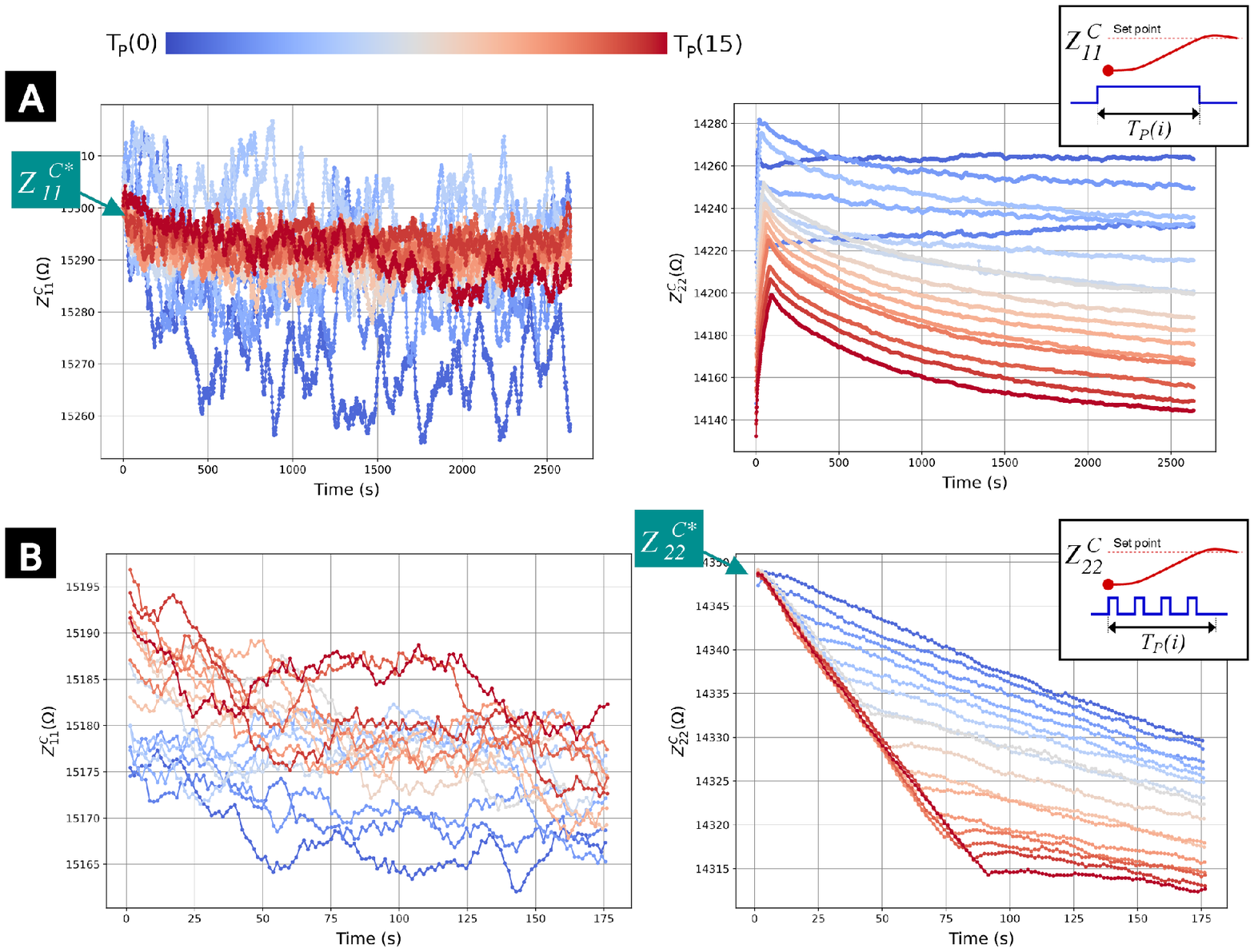}
\caption{\textbf{A} Measured storage capacity obtained by starting from an initial condition $Z_{11}^{C*}$\,=\,15300\,$\Omega$ for the same pulse duration given in Fig.~2A (main manuscript), with a total test time of 2500\,s. \textbf{B} Memorization capability obtained using repeated short duration pulses and resetting the impedance at every iteration to $Z_{22}^{C*}$\,=\,14338\,$\Omega$.}\label{figs6}
\end{figure}
\FloatBarrier

The plots of Fig.~\ref{figs6}A show information storage results for more than 2500\,s obtained by applying the same signals given in Fig.~2A of the main manuscript, starting from an impedance \linebreak set point $Z_{11}^{C*}$\,=\,15300 and by resetting it at every iteration (blue $T_P$\,=\,4\,s, towards red $T_P$\,=\,64\,s). Interestingly, while the effect of the pulses is not appreciable at port one, at port two we can observe that the final impedance values are separated from each other and their separation is maintained even for such long duration.
The graphs of Fig.~\ref{figs6}B show the possibility of carefully mapping pulse duration $T_P$, thus increasing resolution towards a full analog memory. In this test, we have applied the same scheme given in Fig.~2A (main manuscript), but with pulses instead of fixed 3.3\,V signals ($T_\mathrm{high}$\,=\,0.25\,s, $T_\mathrm{low}$\,=\,0.75\,s, for a duty cycle of 25\%, see detail in the plots). The use of these pulses permits the observation of fine variations of storage capacity as information is encoded in the number of pulses where a higher number indicates a higher information value. In this case, programming occurs with a lower \emph{energy} compared to the previous case. Here, however, the obtained impedance values decrease with time but the curves are still separated during their evolution. 

As reported in Fig.~2 (main manuscript) and Fig.~\ref{figs6}, depending on the bias setting of the colloid different numerical values and trends can be achieved. This implicitly demonstrates that the system provides a long term memorization feature that stores the previous stimulation history. Analog memories in general are not a new concept [see \cite{bib71}], but can open the way, similarly to memristors, to unexplored computation capabilities if paired with modern neural networks and machine learning. 
\newpage

\section{Dataset and Differentiation}
\label{Sec9}

\begin{figure}[hbt!]%
\centering
\includegraphics[width=0.75\textwidth]{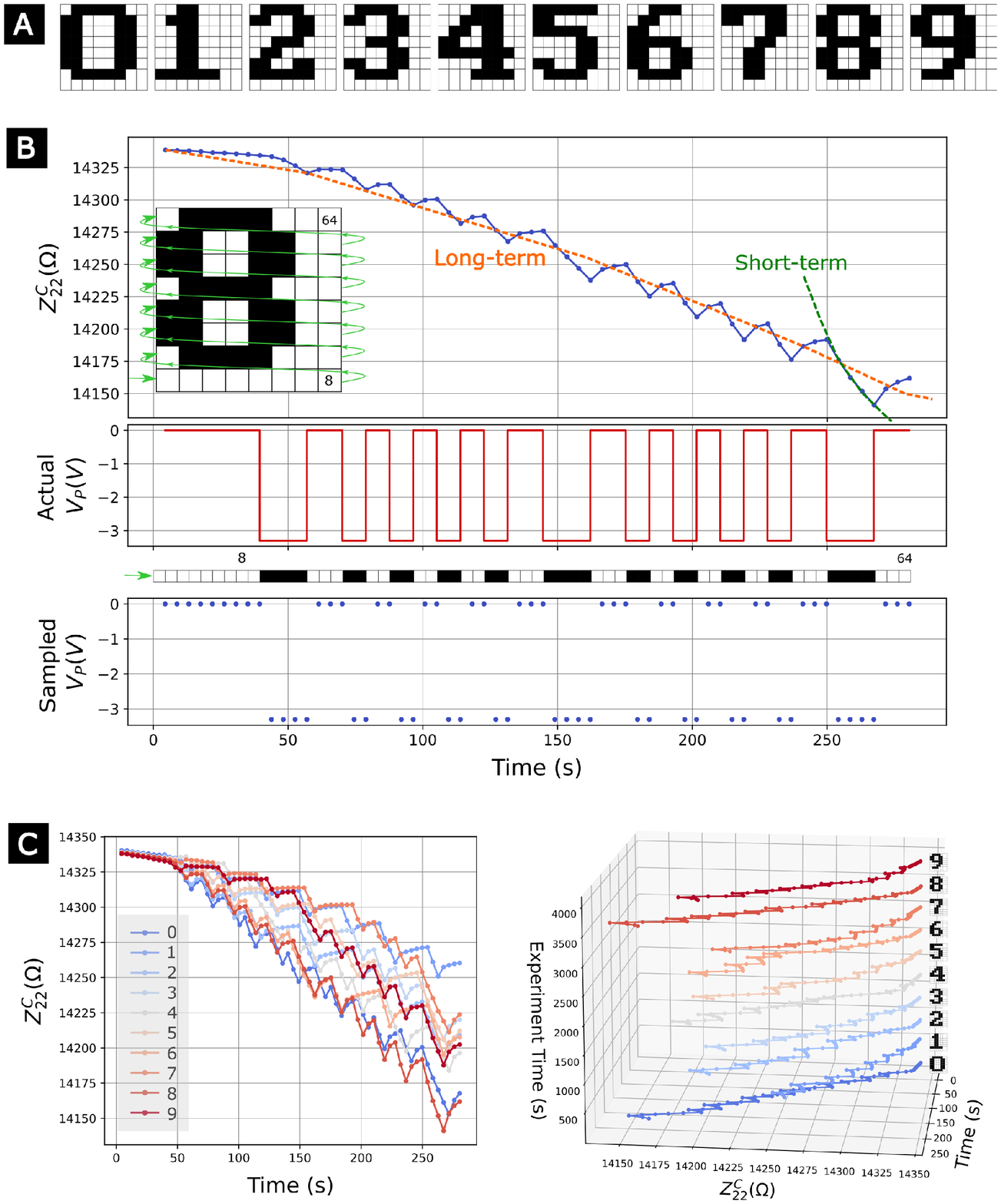}
\caption{\textbf{A} Dataset of ten different 8\,$\times$\,8 pixel images (black and white), each one identifying a digit from {\tt 0} to {\tt 9}. \textbf{B} Measured $Z_{22}^C$ during serialization of digit {\tt 8} using constant weighting, with detail on short and long-term information storage features. \textbf{C} Measured pixel array differentiation assuming constant time-weighting $w_i$\,=\,$w$\,=\,4\,s for the digit dataset (single measurement run). The values obtained at the end of the serialization are all different and depend on both the input sequence and the previous history. Within all the serialization curves, we can appreciate local variations attributable to a short-term memory effect, while final values include both long and short-term memory contributions.} \label{figs5}
\end{figure}
\FloatBarrier

We have investigated the possibility of classifying the different digits assuming constant delay weights $w$, to verify that the colloid operates similarly to a solid state memristive device featuring short-term plasticity, and therefore, thanks to its non-linear lossy integration, differentiate spike sequences \cite{bib67}. To verify differentiation, we have applied a constant interval of 4\,s for all pixels (that can be 1, i.e., -3.3\,V or 0, i.e., 0\,V) of all digits (see dataset in Fig.~\ref{figs5}A). We have run measurements by serializing all digits from 0 to 9 and applying them to the liquid in sequence. 

Fig.~\ref{figs5}B shows an example measurement for digit {\tt 8} during serialization with detail on the pixel values. Our measurement script saves $V_P$ at the end of each stimulation and thus we provided both sampled data and its corresponding zero-order hold interpolation, that identifies the actual voltage across the colloid. From the measurements of $Z_{22}^C$ we can clearly identify two memory contributions, one long-term and another short-term. The short-term contribution fades out rapidly as stimulation is interrupted (for black pixels, -3.3\,V).

Fig.~\ref{figs5}C shows the measurement results for all the digits. The values obtained at the end of each sequence are all different, demonstrating a differentiation, similarly to the results obtained on memristors matrices \cite{bib68}. Observe how the slope of the decreasing curves (that is while a black pixel is applied to the liquid), varies across the serialized digits, to show, similarly to Fig.~\ref{figs5}B an impact of both short and long-term plasticity.
\newpage

\section{Progressive Adaptation}
\label{Sec5}

\begin{figure}[hbt!]%
\centering
\includegraphics[width=0.9\textwidth]{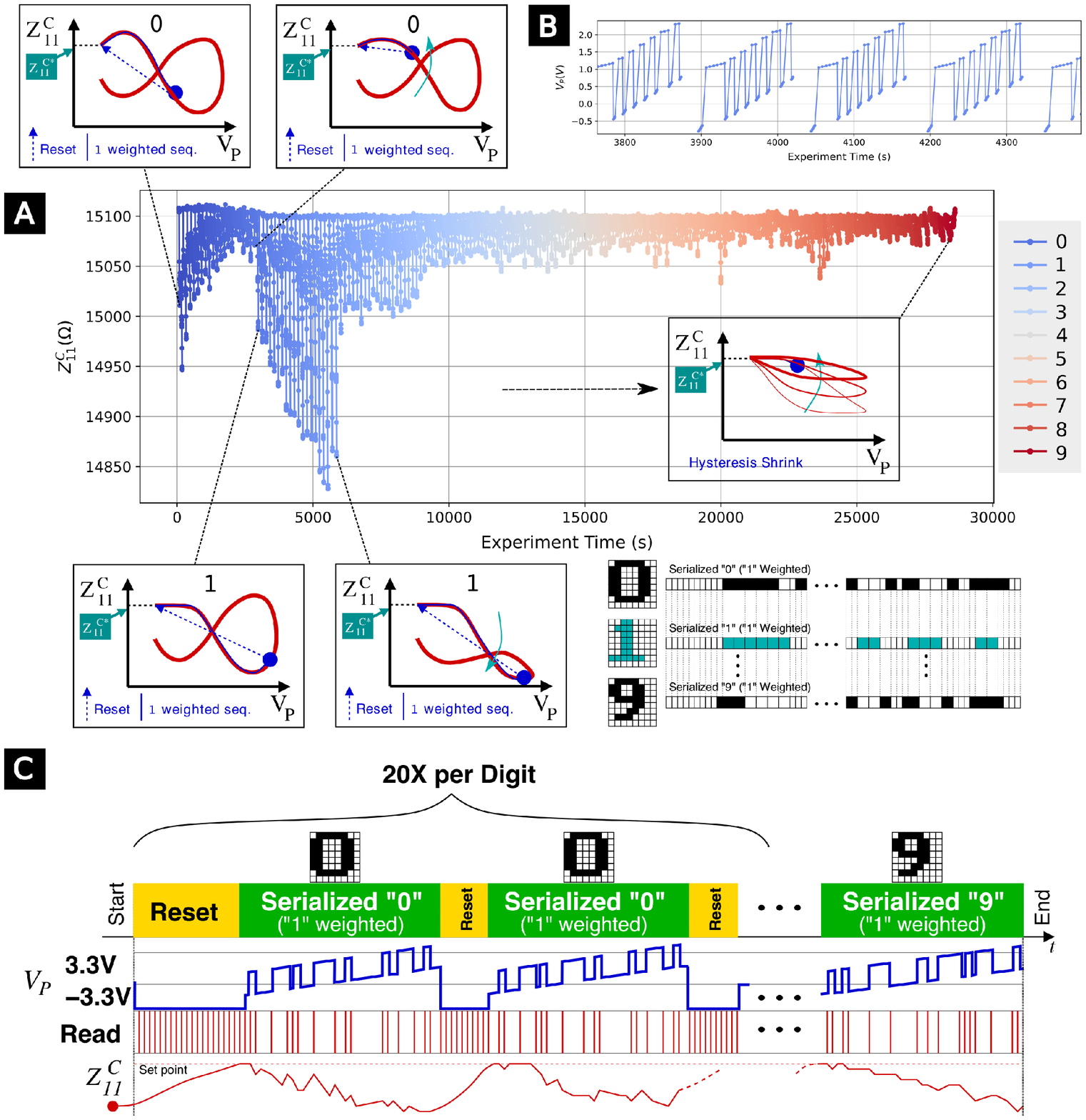}
\caption{\textbf{A} Measured progressive adaptation obtained by running across the hysteresis curve given in Fig.~1C and by repeating the digits 20 times each. The initial impedance set is $Z_{11}^C$\,=\,15100\,$\Omega$. When the correct digit matches weighting (which is constant across all digits, here for digit {\tt 1}), impedance decreases significantly. \textbf{B} Applied voltage across the colloid with progressively increasing bias to run across the hysteresis curve. \textbf{C} Corresponding measurement scheme for the test.}\label{figs2}
\end{figure}
\FloatBarrier

Among the extensive tests we performed on the colloid, we also investigated the possibility of exploiting the pinched hysteresis to demonstrate progressive adaptation to a training sequence, in a similar way as a 'learning' mechanism.
In our in-memory computing scheme, weighting can be applied also with a non-zero offset. Assuming a pinched hysteresis, an offset stimulation is useful to run across it, and depending on the overall duration of the voltage applied, we demonstrate that the liquid can dynamically latch towards low or high impedance values. Fig.~\ref{figs2}A shows measurement results in which we have applied repeatedly and sequentially the digits (20 times each) given a weighted sequence for {\tt 1}. Stimulation runs across the pinched hysteresis loop deeper when {\tt 1} is applied with respect to all the other digits, so to lead impedance dynamically decrease of a larger magnitude. We have reset the impedance to $Z_{11}^{C*}$\,=\,15100\,$\Omega$, exploiting the equilibrium conditions of the material immediately before our hysteresis tests given in Fig.~1C (main manuscript). Compared to the measurements given in Fig.~3 (main manuscript), the FF is here in the conditions such as the voltage values are exchanged, that is obtain a $Z_{11}^{C}$ impedance decrease a positive $V_P$ is required, and consequently a negative one is required to reset it. 
To run across the pinched hysteresis, as shown in Fig.~\ref{figs2}B we have applied a progressive offset to the same weight sequence of the measurements of Fig.~3. 
The offset function is all equal for all digits, and progressively increases as the serial pixel is streamed according to the sequence $K\left(1-\frac{i}{32}\right)$, where $K$ is negative, and $i$ is the pixel number in the range 0--63.
As shown in Fig.~\ref{figs2}C, to permit the establishment of an internal equilibrium and not perturb the colloid status, we have not performed a hard reset with a high 10\,V voltage, rather we have reset it with the same magnitude voltage of 3.3\,V used for serialization. As exemplified qualitatively in the hysteresis diagram details, this measurement comprises mostly all the effects we have observed during our study. In particular, hysteresis progressively evolves depending on the previous history to increase the impedance dynamics for digit {\tt 1} and terminate with an almost full zeroing for digit {\tt 9}, where the pinch disappears.
The impedance dynamics of the colloid then progressively latches towards large variations for {\tt 1} and progressively reduces for the remainder digits.
These results demonstrate the possibility of a dynamical latching using a FF in a similar way it can be done using memristors. Moreover, it demonstrates that a `learning' mechanism is possible.
\newpage

\section{Dynamics Reduction}
\label{Sec6}
\label{sec:limitcycle}

\begin{figure}[hbt!]%
\centering
\includegraphics[width=0.8\textwidth]{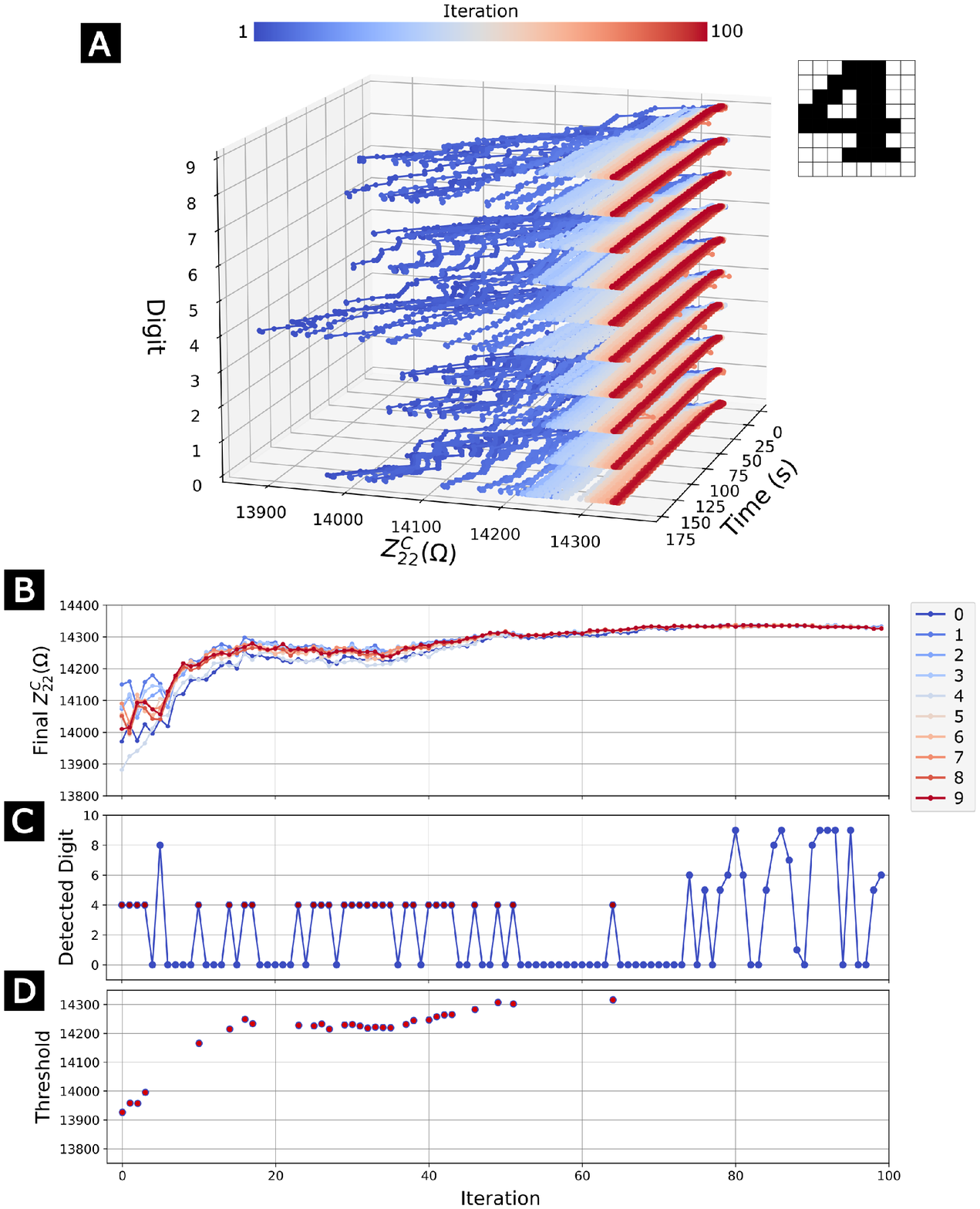}
\caption{\textbf{A} Measured reduction of the FF dynamics after repeating 100 times the classification measurement with weighting for digit {\tt 4}, and initial set point $Z_{22}^C$\,=\,14338\,$\Omega$. The impedance of the system evolves to shrink around a non-dynamical fixed point, but with minimal dynamical features (limit cycle). \textbf{B} Final value of $Z_{22}^C$ after the application of all pixels versus iteration. \textbf{C} Detected digits versus number of iterations. \textbf{D} Corresponding threshold for the detection of {\tt 4}. }\label{figs1}
\end{figure}
\FloatBarrier

In the in-memory classification experiments of Fig.~3 (main manuscript) we have correctly assumed that during the tests
the status of the liquid was not significantly changing. This hypothesis was verified because the tests lasted less than one hour each (see the experimental time in the measurement results), for an overall duration of approximately 5\,h. However, in general,  
we have observed that after applying the serial sequences for days, the resetting of the material to the same condition at every test leads to a dynamics reduction towards the impedance set point itself. This result is consistent with the hysteresis plots of Fig.~1 (main manuscript): assuming a periodical stimulation, its effect on impedance varies over time, leading, in the ultra-long term, to a shrinking of the hysteresis curve of $Z_{22}^C$.
%This fact is evident in these particular classification measurements, because impedance reset of the liquid on one port is achieved by applying a higher and invariant 10\,V pulse compared to the -3.3\,V level of pixel serialization, until the initial impedance set point is reached. 

Fig.~\ref{figs1}A shows superposed $Z_{22}^C$ curves 
for each digit, when the pattern classification test is repeated in sequence for 100 times with weighting for digit {\tt 4} and overall measurement time of 3 days. The dynamics of the impedance patterns obtained during the measurements tend to collapse to a straight line superposed to the initial impedance set point, with reduced dynamical behaviour.
Fig.~\ref{figs1}B shows the final $Z_{22}^C$ values after the streaming of all pixels as a function of iteration. We can define a detection threshold as the average value between the lowest and the immediately preceding value of $Z_{22}^C$ across all digits at a given iteration. Due to dynamics reduction, the in-memory classification scheme fails as the lowest impedance values do not always correspond to {\tt 4}, notwithstanding weighting. The detected digits shown in Fig.~\ref{figs1}C, confirm that digit {\tt 4} leads to the lowest impedance value 29 times out of 100, for an accuracy of 29\%. The expected detection threshold for {\tt 4}, as shown in Fig.~\ref{figs1}D, follows the general compression trend of the liquid as the classification is progressively repeated. 

Consequently, within the given experiment time, we can conclude that the system evolves towards a limit cycle, making in-memory computing with the previously defined scheme ineffective. We observed that this behaviour is induced by the reset phase of the impedance achieved in the same increasing direction for all interactions, which, \emph{inter alia}, is performed at higher voltage (i.e., 10\,V) compared to those used during the serialization (i.e., -3.3/3.3\,V). The behaviour of the liquid, however, is reversible. After applying a fixed stimulation of -10\,V for a sufficiently long time (on the order of minutes) the dynamical behaviour of the liquid is increased and restored. Moreover, the liquid dynamics can be also restored by repeatedly applying fast and zero mean hysteresis sweeps at higher voltage (e.g. a factor two faster compared to the reported results of Fig.~1C, main manuscript). Another possibility is to pose $V_P$\,=\,0\,V and wait for a sufficiently long time to enforce the system towards an internally defined equilibrium, thus leaving the liquid to self-adjust. We have observed that after such restoring, although the initial impedance values can significantly vary compared to the preceding epoch, the dynamical behaviour is still restored. After this restoring process, the impedance values can be successfully brought back to the original values by applying DC voltage.

Given that such locking phenomenon occurs after many repeated classification cycles and in general after long time, provided that the computation time is reasonably small, the liquid can be still effectively used to run computation, even assuming the above dynamics reduction. This limit cycle (that can be considered as a long-term effect) fades out, but slower than the computation time of the above experiments thanks to the relatively low voltages used in the serialization process. We have observed faster dynamics reduction with higher stimulation voltage. 

\newpage

\section{Solid-state Memristors vs. Ferrofluids}
\label{Sec2}
Two features of FF that can be different compared with typical solid-state memristors are, i) the application of DC bias multiple times to change the material state and ii) speed of the inference operation.
Most memristor-based neural networks are inferred at biological speed, i.e., in a millisecond-range time constant. With oscillatory neurons, however, one can vary this time constant by several orders of magnitude, for instance between 1\,$\mu$s and 100\,s or even 1\,ks. On the other hand, the FF needs to be read at RF-level (using an AC signal), unlike memristors, to the best of our knowledge.
Contrarily to memristors where the impedance observed is mono-dimensional (applied voltage, read current), here the status of the liquid is observable across all four impedance parameters, thus potentially allowing for higher dimensional interpretations. Another interesting point, regards the application of programming biases repeated times. This feature is valid for both FF and some memristors. HfO$_{2}$-based memristors, for instance, need this kind of multiple biasing approach to witness a change of state. This is mainly due to their conductive filament-based switching mechanism they provide \cite{bib75}.
\newpage

\section{Sensitivity to Initial Conditions}
\label{Sec7}

\begin{figure}[hbt!]%
\centering
\includegraphics[width=0.9\textwidth]{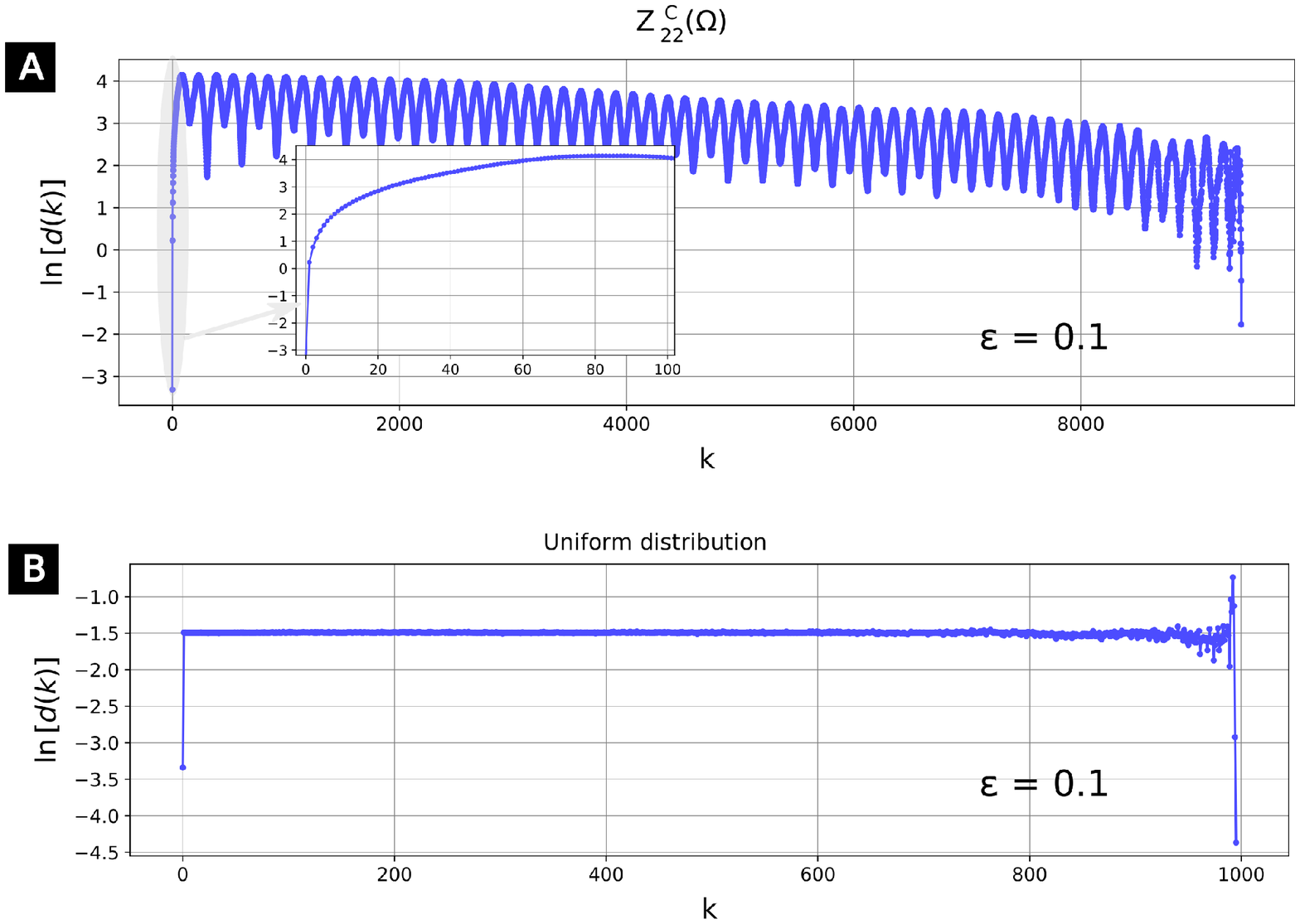}
\caption{\textbf{A} Computed series $\ln \left[ d(k) \right]$ associated to the $Z_{22}^C$ time-series of our hysteresis test in Fig.~1C ($\epsilon$\,=\,0.1). The slope at the beginning of the test is positive and therefore exists a linear fit with positive slope that is a reasonable estimation of the Lyapunov exponent. \textbf{B} The same computed series of A, but for a uniform distributed function. The trend is flat, therefore $d(k)$ does not depend on $k$, confirming stochasticity.}\label{figs3}
\end{figure}
\FloatBarrier

A defining feature of a chaotic deterministic systems is its high sensitivity to initial conditions. To check whether our system provides this feature we have extracted the $Z_{22}^C$ time-series given in Fig.~1C (for the complete experiment) and we have applied the algorithm of \cite{bib74} which provides a means to estimate the Lyapunov exponent, based on the quantity $d(k)=\lvert T_{i+k}-T_{j+k}\rvert$, where $T$ is the time-series. Fig.~\ref{figs3}A shows the results of the application of the algorithm with an initial diameter bound of $\epsilon$\,=\,0.1. The curve exhibits periodicity because the data comprises all the loops depicted in the hysteresis. In contrast to the uniform distribution case (which provides a flat behavior), the term $\ln \left[d(k)\right]$, at the beginning of the experiment, shows an increasing trend with $k$, suggesting that exists a linear fit with positive slope, which is a good estimation of the Lyapunov exponent. A positive Lyapunov exponent identifies chaotic character of the time-series and hence a high sensitivity to initial conditions, as well as a system where the entropy creation rate is positive \cite{chaos}.
\newpage

\section{Chaotic Non-Equilibrium}
\label{Sec8}

\begin{figure}[hbt!]%
\centering
\includegraphics[width=0.9\textwidth]{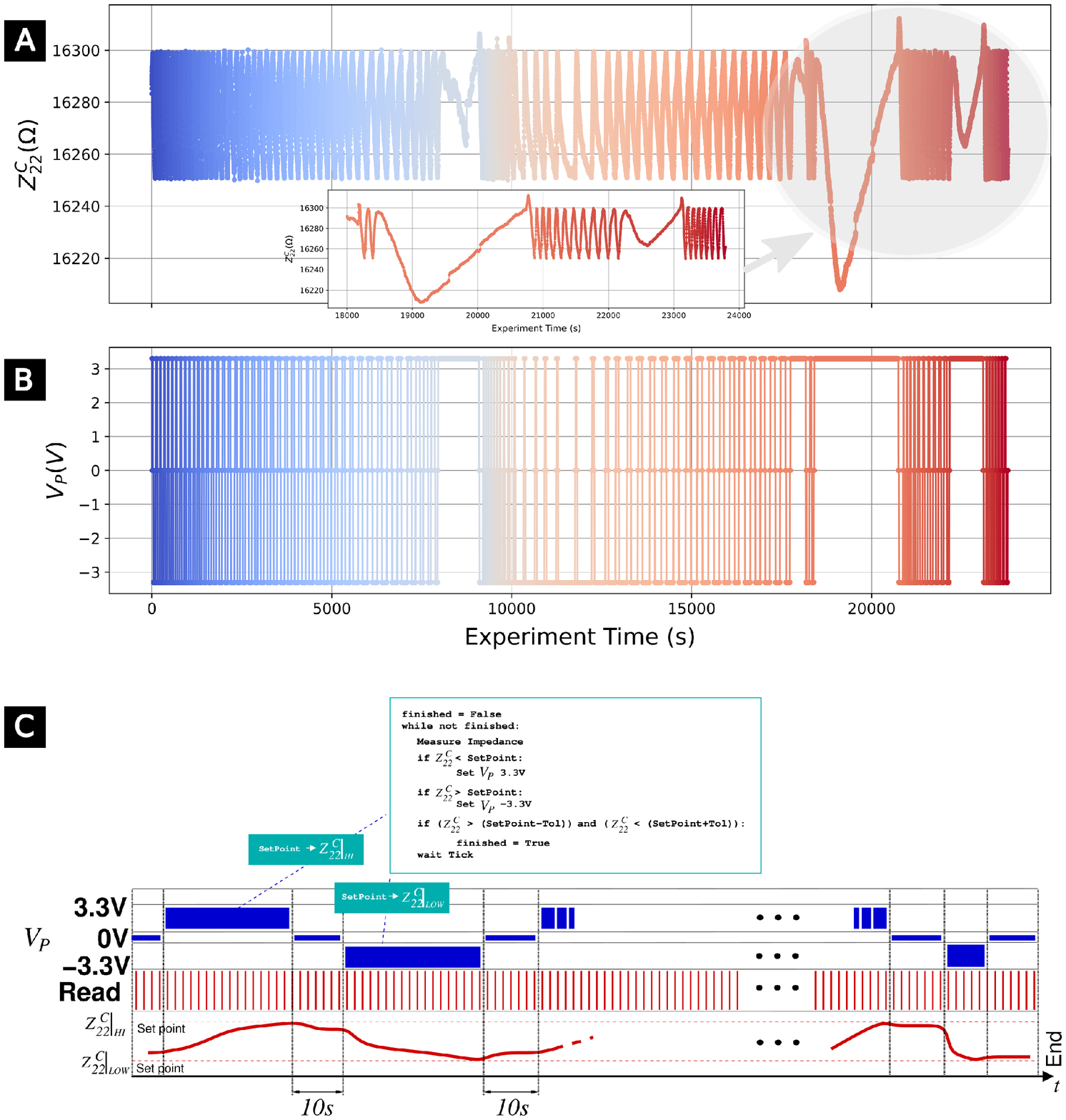}
\caption{\textbf{A} Measured $Z_{22}^C$ impedance during a repeated test in which the liquid, thanks to the depicted control algorithm ({\tt Tol}\,=\,1, {\tt Tick}\,$\sim$\,0.7\,s) is brought to two particular impedance values, $Z_{22}^C\rvert_\mathrm{LOW}$\,=\,16250\,$\Omega$ and $Z_{22}^C\rvert_\mathrm{HIGH}$\,=\,16300\,$\Omega$. The plot shows chaos in the oscillatory regimes. \textbf{B} Applied voltage $V_P$ during the experiment. \textbf{C} Summary of the measurement scheme depicting where the control algorithm intervenes to bias the material to the two particular impedance points.}\label{figs4}
\end{figure}
\FloatBarrier

In view of the reported properties of the FF, and to confirm the obtained results, we have run an experiment to further demonstrate that the system is chaotic and features a chaotic non-equilibrium, which could also explain the evolution over time of the hysteresis curves. Fig.~\ref{figs4} shows the results of this experiment. The test consists of the multiple setting of two impedance values, with a control algorithm that enforces $V_P$\,=\,3.3\,V or -3.3\,V across the liquid to set the impedance state to $Z_{22}^C\rvert_\mathrm{HIGH}$ and $Z_{22}^C\rvert_\mathrm{LOW}$, respectively. In between stimuli, we applied pauses of 10\,s where voltage $V_P$ is zero (see the measurement scheme of Fig.~\ref{figs4}C). The algorithm applies a fixed voltage until impedance reaches a predefined set point, but the reached impedance value is not exactly the same as the previous iteration, because the measurement accuracy of the set-up is finite. This way, initial conditions of the next iteration are slightly perturbed, and we can then determine if the response of the liquid varies over time. Results in Fig.~\ref{figs4}A depict the switching between various oscillatory states with different rates across $Z_{22}^C$. In particular, see Fig.~\ref{figs4}B, when the voltage $V_P$ applied across the material is fixed (in this example 3.3\,V but our findings show that it occurs also for -3.3\,V biases), the impedance variation that is expected to be monotonically increasing, instead varies with a more complex trend, to separate the oscillatory states in subgroups. The presence of multiple oscillatory regimes can be assumed to pertain to metastable attractors, while overall the resulting complex oscillation can be assumed as chaotic.

Being the FF a chaotic dynamical system from a DC/RF viewpoint, it can always be seen as an Analog Recurrent Neural Network (ARNN) \cite{bibmst}, therefore as an ensemble of nodes of a reservoir. Given such a high degree of complexity the system exhibits, we conclude that the colloid can be considered not only a liquid synapse but an ensemble of more complex nodes, all inter-related: a liquid network. Considering the measurements of Fig.~\ref{figs4}, a possible explanation of the phenomena, can be traced back to a network of oscillators. Indeed, each particle of the FF can be seen as a single superparamagnetic oscillator that is coupled via dipolar magnetic fields to those surrounding it. Considering the size of the particles in the FF (nanometric) and the material itself, magnetite, the spin resonance frequency is in the order of 1--3\,GHz, within the range of the conducted experiments. Hence, the application of the RF field to readout the status of the material surely comprises the effects of the ferromagnetic resonance of these oscillators. The effect of the DC stimulus could be explained by the application of a torque on the oscillators with respect to their natural position, generated on each local spin by the magnetic field. 
\newpage

\section{Physical Reservoir Computing -- Readout Neural Network Training and Notes}
\label{Sec10}

\begin{figure}[hbt!]%
\centering
\includegraphics[width=0.9\textwidth]{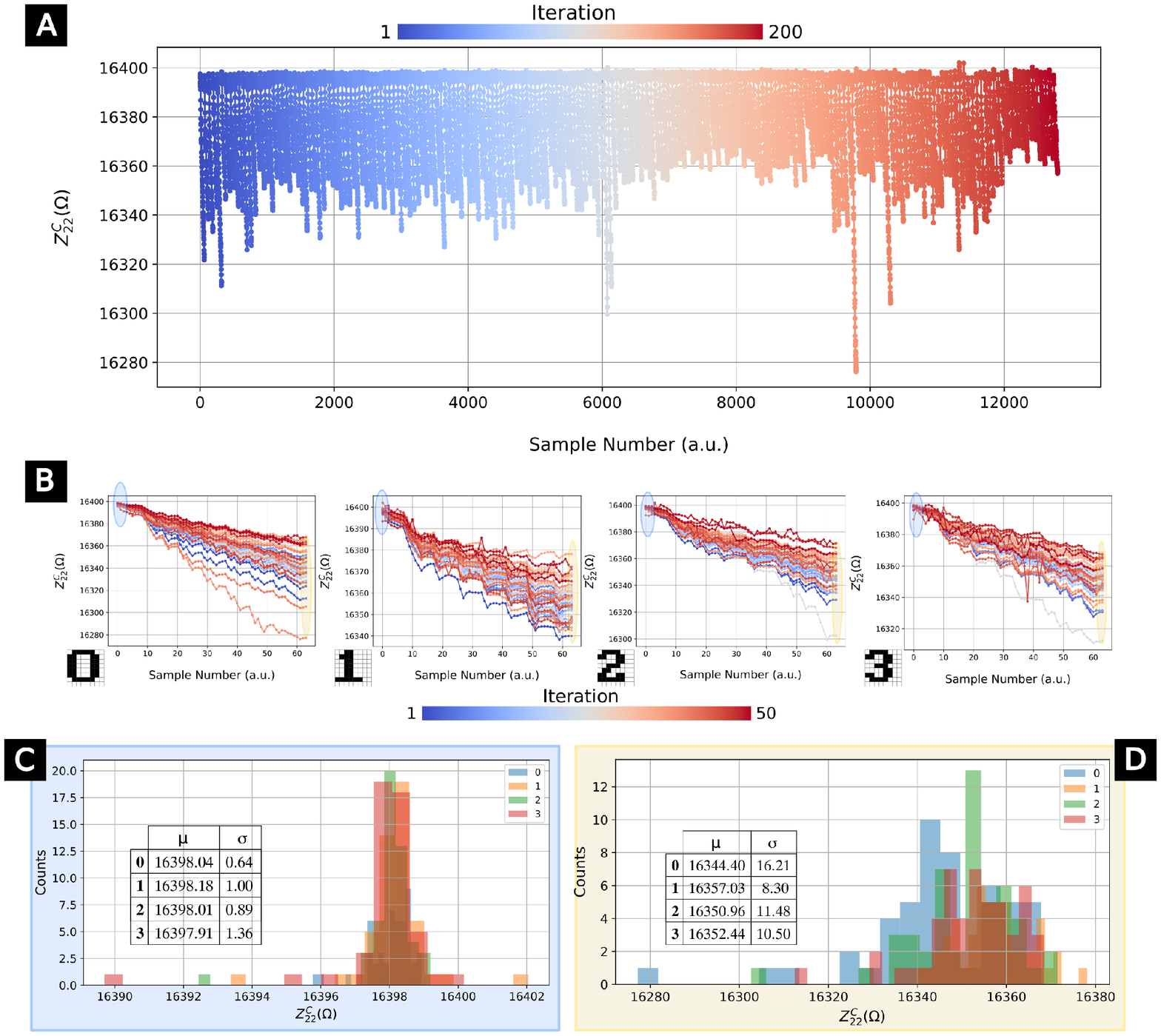}
\caption{\textbf{A} Complete $Z_{22}^C$ values corresponding to 50 serializations of the digits {\tt 0}--{\tt 3} (overall 4\,$\times$\,50), for the training of the readout neural network used in PRC. Notwithstanding the 'reset' condition maintains dynamical properties of the material alive, outputs have variability. \textbf{B} Superposed $Z_{22}^C$ values for each digit for all the 50 iterations. \textbf{C} and \textbf{D} Histograms of the values of $Z_{22}^C$ for the four digits at the first iteration (left) and at the last iteration (right).} \label{figs7}
\end{figure}
\FloatBarrier

Fig.~\ref{figs7}A shows the complete waveform of $Z_{22}^C$ used to train the neural network depicted in Fig.~4 (main manuscript), later used to perform real-time detection. As shown in Fig.~\ref{figs7}A, the periodical streaming of the digits (repeated {\tt 0}--{\tt 3} sequence) leads to a high variability in the impedance values, even with a 'reset' sequence that maintains the liquid in a dynamical condition. This notwithstanding, the detail of each digit given in Fig.~\ref{figs7}B shows the successful set of the initial impedance value thanks to the control loop that here is set to an unilateral tolerance of 2.5\,$\Omega$.
A detail in the distribution of impedance values is given in Fig.~\ref{figs7}C--D that show histogram bars of both first and last $Z_{22}^C$ values for all digits, with normal distribution fitting (mean and standard deviation, $\mu$ and $\sigma$).

The 'reset' sequence, used to counterbalance the chaotic features of the FF, enables the application of a normalization layer as first element in the neural network.
Observe that the 'reset' sequence can be considered as a means to enforce the Echo State Property (ESP) \cite{esp}, a key requirement to achieve training using only the output of the reservoir. The FF provides both long-term and short-term information storage and a required condition for ESP is that the effect of initial condition should vanish as time passes. Without a 'reset' sequence, this property is definitely not verified for our reservoir assuming the same order of magnitude time required for digit classification. Waiting for a complete fade out of the effect of the inputs would make RC unreasonably slow.

In memristors PRCs, depending on the specific computing application, the readout NN can be for instance a Convolutional Neural Network (CNN) to account for the time dependency of subsequent samples in real-time detection of firing patterns, or a fully connected NN for pattern recognition \cite{bib67}. Here, the monothonically decreasing curve of impedance ensured by the 'reset' sequence, permits a readout network implementation based on fully connected {\tt Dense} layers. The readout layer given in Fig.~4 (main manuscript), provides excellent training results, i.e., 3\,$\times 10^{-4}$ loss for normalized outputs ({\tt 0}--{\tt 3}\,$\rightarrow$\,0--1) on 2000 epochs, resulting in a 0.001 Root-Mean-Square Error (RMSE). Hence, this NN permits inference on new real-time data by enabling an easy counterbalancing of the chaotic variations of our reservoir. However, the definition of RC typically refers to a single output layer to map the higher dimensional space projected by the reservoir [that internally implements a Recurrent Neural Network (RNN)] to the desired features. This corresponds to a single fully connected neuron in our specific case, that implements our single dimensional digit mapping {\tt 0}--{\tt 3} (normalized between 0 and 1). 
Although with lower training accuracy compared to the previous case, by using a single {\tt Dense} layer (sigmoid activation function), input normalization and the dataset given in Fig.~\ref{figs7}, we obtained a loss of 0.044 on 25000 epochs (0.062 RMSE). By using two {\tt Dense} layers, the first with four neurons, and the second with one neuron, with input normalization we obtained 0.015 loss on 25000 epochs for a 0.026 RMSE. These values are still reasonable to implement PRC with a lower complexity readout NN.

%\bibliography{scibib}% common bib file
%\bibliographystyle{Science}
